\documentclass[12pt]{article}
\usepackage{graphicx}
\def\pw#1{^{#1}}

\def\beeq{\begin{eqnarray}} \def\eeeq{\end{eqnarray}}
\newcommand\mysection{\setcounter{equation}{0}\section}
\renewcommand{\theequation}{\thesection.\arabic{equation}}
\newcounter{hran} \renewcommand{\thehran}{\thesection.\arabic{hran}}

\def\bmini{\setcounter{hran}{\value{equation}}
  \refstepcounter{hran}\setcounter{equation}{0}
  \renewcommand{\theequation}{\thehran\alph{equation}}\begin{eqnarray}}

\def\bminiG#1{\setcounter{hran}{\value{equation}}
\refstepcounter{hran}\setcounter{equation}{-1}
\renewcommand{\theequation}{\thehran\alph{equation}}
\refstepcounter{equation}\label{#1}\begin{eqnarray}}

\def\emini{\end{eqnarray}\relax\setcounter{equation}{\value{hran}}\renewcommand{\theequation}{\thesection.\arabic{equation}}}

\def\fun#1#2{\lower3.6pt\vbox{\baselineskip0pt\lineskip.9pt
  \ialign{$\mathsurround=0pt#1\hfil##\hfil$\crcr#2\crcr\sim\crcr}}}

\def\ben{\begin{enumerate}}  \def\een{\end{enumerate}}
\def\bit{\begin{itemize}}    \def\eit{\end{itemize}}
\def\beq{\begin{equation}}   \def\eeq{\end{equation}}
\def\bea{\begin{eqnarray}}  \def\eea{\end{eqnarray}}
\def\nn{\nonumber}
\def\noi{\noindent}

\def\kp{\relax\ifmmode{k_\perp}\else{$k_\perp${ }}\fi}
\def\kps{\relax\ifmmode{k_\perp\pw2}\else{$k_\perp\pw2${ }}\fi}
\def \as{\relax\ifmmode\alpha_s\else{$\alpha_s${ }}\fi}

\newskip\humongous \humongous=0pt plus 1000pt minus 1000pt
\def\caja{\mathsurround=0pt}

\newif\ifdtup

\def\eqal2#1{\,\vcenter{\openup1\jot
\caja   \ialign{\strut \hfil$\displaystyle{##}$&\hfil$
\displaystyle{{}##}$\hfil &$
\displaystyle{{}##}$\hfil\crcr#1\crcr}}\,}

\def\lsim{\raise0.3ex\hbox{$<$\kern-0.75em\raise-1.1ex\hbox{$\sim$}}}
\def\gsim{\raise0.3ex\hbox{$>$\kern-0.75em\raise-1.1ex\hbox{$\sim$}}}
 \def\cite#1{[\ref{#1}]}
 \def\citd#1#2{[\ref{#1},\ref{#2}]}
 
 \def\citm#1#2{[\ref{#1}--\ref{#2}]}

\begin{document}
\begin{center}
{\bf THE RUNNING GAUGE COUPLING IN THE} \\
{\bf EXACT RENORMALIZATION GROUP APPROACH} \\
\vspace{2 truecm}
{\bf Ulrich Ellwanger}\footnote{email : ellwange@qcd.th.u-psud.fr}\\
Laboratoire de Physique Th\'eorique et Hautes
Energies\footnote{Laboratoire associ\'e au Centre National de la Recherche Scientifique
(URA D0063)}\\    Universit\'e de Paris XI, Centre d'Orsay, 91405
Orsay Cedex, France\\  
\end{center}
\vspace{3 truecm}
\noi {\bf Abstract} \par

We discuss the perturbative running Yang-Mills coupling constant in the Wilsonian exact
renormalization group approach, and compare it to the running coupling in the more
conventional $\overline{MS}$ scheme. The exact renormalization group
approach corresponds to a particular renormalization scheme, and we relate
explicitly the corresponding $\Lambda$ pa\-ra\-me\-ters. The unambiguous definition of an
exact renormalization group scheme requires, however, the use of a one-loop improved high
energy effective action. \\

\vspace{1 truecm} 

\noi LPTHE Orsay 97-02 \\ 
\noi January 1997 \\
 
\newpage
\pagestyle{plain}
\mysection{Introduction}

Some time ago the exact renormalization group equations (ERGEs) of Wegner and Wilson
\cite{1r} have been formulated in continuum quantum field theory by Polchinski \cite{2r}.
The original motivation was the simplification of the proof of renormalizability, and a
better understanding of its meaning in cutoff field theories. Since then it has become
clear that the ERGEs also provide a computational tool for the determination of a ``low
energy'' effective action in terms of a ``bare'' (high energy) action~: they describe
the continuous evolution of effective Lagrangians (or effective actions) with a scale
(or infrared cutoff) $k$, which can be integrated from a cutoff scale $^{-}\hskip - 2.5
truemm{k}$ (sometimes denoted by $\Lambda$ elsewhere) down to $k = 0$. In contrast to
standard renormalization group equations the ERGEs describe this evolution including all
irrelevant couplings, or higher dimensional operators, and are exact in spite of the
appearance of one-loop diagrams only. They correspond, however, to an infinite system of
coupled differential equations, which has to be approximated for practical purposes. The
advantage of the method is, on the other hand, its nonperturbative nature and its large 
flexibility, which allows for many different approximation schemes (which may vary with 
the scale $k$). \par

One can, e.g., expand the effective action in powers of momenta or derivatives, keeping
all powers of the involved fields (see \cite{3r} for some early literature). This
approach shares some features with the formulation of the theory on a finite size
lattice, since on a lattice of size $N$  only derivatives of the order $N - 1$ can be
defined. 
Alternatively, one can expand the effective action in powers of fields, keeping all
powers of the momenta \cite{4r}. This kind of expansion is familiar from the truncation
of the infinite series of Schwinger-Dyson equations. \par

The application of ERGEs to gauge theories has to surmount the problem that the
intermediate infrared cutoff generally breaks the gauge invariance \citm{5r}{9r}.
One way to solve this problem is the use of modified Slavnov-Taylor identities
\citm{7r}{9r}, which impose ``fine tuning conditions'' on those couplings in the
effective action at scales $k \not= 0$, which break gauge or BRST invariance. These
``fine tuning conditions'' ensure the BRST invariance of the full effective action
for $k \to 0$. \par

Recently the first steps in the application of the ERGE approach to QCD were
undertaken \cite{9r}. In \cite{9r} the ERGEs for the gluon and ghost propagators in
pure SU(3) Yang-Mills theory were integrated within an approximation, which was based
on an expansion of the ERGEs in powers of fields and, at the same time, imposed the
modified Slavnov-Taylor identities on the effective action. A certain combination of
the gluon and ghost propagator functions can be identified with the part of the heavy
quark potential (in the quenched approximation), which is induced by ``dressed'' one
gluon exchange (with a dressed gluon propagator and a dressed quark-gluon vertex~;
this combination is invariant under gluon field redefinitions). \par

The result of the numerical integration of the ERGEs was a form of the heavy quark
potential $V(q^2)$, which has the perturbative one-loop form for $q^2 \to \infty$,
but which shows a confining behaviour (like $\displaystyle{{1 \over q^4}}$) for small
$q^2$. Actually the approximation employed in \cite{9r} ceased to be reliable for
$q^2 \to 0$, but already within the trustworthy range of $q^2$ the confining
behaviour is evident, and it happens to be quite well described by a form proposed
by Richardson \cite{10r}. \par

 From this result, one can define a dimensionful non-perturbative quantity like the
slope of the potential in ordinary space, or a string tension $\sigma$. On the
other hand, the only input of the calculation was a ``bare'' scale invariant
action~; the fact, that this input action plays the role of an effective action at
a ``large'' scale $^{-}\hskip - 2.5 truemm{k}$, is only implicit in the choice of
a small bare gauge coupling constant. Clearly, in pure Yang Mills theory one can
calculate only dimensionless quantities like the ratio $\sigma$/$^{-}\hskip - 2.5
truemm{k}$, as a function of the ``bare'' input gauge coupling constant. \par

It is fairly easy to see (by reintroducing the Planck constant $\hbar$) that the
iterative solution of the ERGEs for the effective action \cite{11r} reproduces
the perturbative series. Thus, for asymptotically free theories, the ``bare''
gauge coupling constant runs with the scale $^{-}\hskip - 2.5 truemm{k}$, for large enough
$^{-}\hskip - 2.5 truemm{k}$, as in perturbation theory. Eventually one would
like to relate it to the running gauge coupling within other perturbative schemes
as the $\overline{MS}$ scheme. This would then allow to relate the
corresponding scales and, finally, to express dimensionful nonperturbative
quantities like $\sigma$ in terms of $\Lambda_{\overline{MS}}$. \par

The aim of the present paper is the derivation of the relation between the running
coupling in the ERGE approach and the $\overline{MS}$ scheme or, equivalently,
the relation between the corresponding $\Lambda$ parameters used to parametrize
the respective scale dependence. This task is thus very similar to the one of
Celmaster and Gonsalves \cite{12r}, who related the running couplings in minimal
and momentum substraction schemes. At the same time we have to clarify, of
course, the meaning of the running coupling constant in context of the Wilsonian
exact renormalization group.

\mysection{The exact renormalization group approach}

The ERGE approach in continuum quantum field theory has already been introduced
and discussed in quite many papers \citm{2r}{11r}, so we will only repeat its
essential features here (restricting ourselves, to this end, to the case of a
single scalar field). \par

To start with, one considers the Euclidean partition function of a theory in the
presence of an infrared cutoff $k$, which is implemented through a modification of
the term quadratic in the field in the bare action $S_0(\varphi)$. The generating 
functional of connected Green functions $G_k(J)$ is thus of the form        

\beq
e^{-{G}_k(J)} = N \int {\cal D}{\varphi}_{Reg} \ e^{-S_0(\varphi) - \Delta S_k
(\varphi ) + (J, \varphi )}
\label{2.1e} 
\eeq

\noi where $(J, \varphi )$ is a short-hand notation for
\beq
(J, \varphi ) \equiv \int {d^4p \over (2 \pi )^4} J(p) \ \varphi(- p) \ \ \ .
\label{2.2e}
\eeq

\noi The index ``Reg'' to the functional integration measure indicates some ultraviolet
regularization (see below), and

\beq
\Delta S_k(\varphi) = {1 \over 2} \left ( \varphi , R_k(p^2) \varphi \right ) 
\label{2.3ea}
\eeq 

\noi serves as the announced infrared cutoff. In the case of a massless field a
convenient choice for the IR cutoff function $R_k(p^2)$ is 

\beq
R_k(p^2) = p^2 {e^{-{p^2 \over k^2}} \over 1 - e^{- {p^2 \over k^2}}}
\label{2.3eb}
\eeq

\noi such that the full propagator

\beq
\left ( p^2 + R_k(p^2) \right )^{-1} = {1 - e^{-{p^2 \over k^2}} \over p^2}
\label{2.4e}
\eeq

\noi is finite for $p^2 \to 0$, and its derivative with respect to $k^2$ vanishes
exponentially for large $p^2$. \par

The ERGE for ${G}_k(J)$ is obtained by differentiating eq. (\ref{2.1e}) with
respect to $k$, and replacing $\varphi$ under the path integral by variations with
respect to the sources. Then one can switch to the effective action $\Gamma_k(\varphi
)$ by a Legendre transform,

\beq
\Gamma_k (\varphi ) = {G}_k(J) + (J, \varphi ) \equiv \widehat{\Gamma}_k(\varphi )
+ \Delta S_k \ \ \ . \label{2.5e}
\eeq

\noi (The present field $\varphi$ is defined, as usual, by the negative of the first
derivative of ${G}_k(J)$ with respect to $J$). The ERGE for $\Gamma_k(\varphi )$
finally assumes the form 

\beq
\partial_k \widehat{\Gamma}_k(\varphi ) = {1 \over 2} \int {d^4q \over (2 \pi )^4}
\partial_k R_k(q^2) \cdot \left ( {\delta^2 \widehat{\Gamma}_k(\varphi ) \over \delta
\varphi (q) \delta \varphi (- q)} + R_k(q^2) \right )^{-1} \ \ \ . \label{2.6e} \eeq

\noi In the case of several fields, as vector fields and ghosts in the case of
Yang-Mills theories, the integral $d^4q$ has simply to be extended by a (super-)trace
over all degrees of freedom, and the inverse on the r.h.s. of (\ref{2.6e}) has to be
replaced by the inverse matrix of second derivatives of $\Gamma_k
+ \Delta S_k$ with respect to all fields \cite{8r}. \par

The physical meaning of the effective action $\Gamma_k(\varphi )$, as defined by eq.
(\ref{2.5e}), is the following~: the coefficients of an expansion of $\Gamma_k(\varphi)$
in powers of $\varphi$ will be called vertex functions, which will in general depend on
the external momenta. These vertex functions include all quantum effects, or all
one-particle irreducible Feynman diagrams, where the internal propagators are
supplemented with an infrared cutoff $k$ as in eq. (\ref{2.4e}). For large $k$, and for
asymptotically free theories, $\Gamma_k(\varphi )$ approaches the classical bare action
of the theory. \par

The full quantum effective action of the theory is obtained from $\Gamma_k(\varphi )$ in
the opposite limit $k \to 0$. Within the ERGE approach it is constructed by
integrating the ERGE (\ref{2.6e}) from some large value of $k$, $k =$ $^{-}\hskip - 2.5
truemm{k}$, down to $k = 0$. The starting point within the ERGE approach is thus an
ansatz for the ``high energy'' effective action $\Gamma_{^{-}\hskip - 2.5
truemm{k}}(\varphi)$ in the case of asymptotic freedom. \par

Universality now tells us that $\Gamma_k(\varphi )$ for $k \to 0$ is independent from
minor changes of $\Gamma_{^{-}\hskip - 2.5 truemm{k}}(\varphi)$ (again for $^{-}\hskip -
2.5 truemm{k}$ large enough)~; therefore, at first sight, we may identify
$\Gamma_{^{-}\hskip - 2.5 truemm{k}}(\varphi)$ with the classical bare action for some
large but finite value of $^{-}\hskip - 2.5 truemm{k}$. (On the other hand, in order to
shorten the domain of $k$ over which eq. (\ref{2.6e}) has to be integrated 
 in the case of practical applications, it may be
useful to use for $\Gamma_{^{-}\hskip - 2.5 truemm{k}}(\varphi)$ a ``one-loop improved''
high energy effective action, see ref. \cite{9r}).
\par

Since, in any case, the starting point within the ERGE approach is a finite ansatz for
a high energy effective action $\Gamma_{^{-}\hskip - 2.5 truemm{k}}(\varphi )$,
problems related to ultraviolet divergences or the need to regularize the theory in the
ultraviolet never arise~: by definition $\Gamma_{^{-}\hskip - 2.5 truemm{k}}(\varphi)$
includes already all quantum effects involving momenta $q^2 \ \gsim \ ^{-}\hskip - 2.5
truemm{k}^2$, but one never has to construct $\Gamma_{^{-}\hskip - 2.5
truemm{k}}(\varphi )$ explicitly in terms of divergent Feynman diagrams and divergent
``bare'' parameters. The construction of $\Gamma_0(\varphi )$ in terms of $\Gamma_{^{-}\hskip - 2.5
truemm{k}}(\varphi )$ by integrating the ERGE (\ref{2.6e}) involves only momenta $q^2$
with $0 \ \lsim \ q^2 \ \lsim \ ^{-}\hskip - 2.5 truemm{k}^2$. \par

It is possible to define running coupling constants from the $k$ dependent vertex
functions at, e.g., vanishing external momenta. ($k$ regularizes infrared divergences
also for exceptional external momenta). From the iterative solution of the ERGE
(\ref{2.6e}) one can obtain the $\beta$ functions for the running coupling constants,
which describe their dependence on $k$. For classically scale invariant theories
(without dimensionful parameters in the bare action) the one and two-loop coefficients
of the $\beta$ functions are scheme independent. The present method defines an ``ERGE
scheme''. (The precise definition will depend on the vertex function used to
define the coupling constant, and the form of the infrared cutoff function $R_k(q^2)$
in (\ref{2.3eb})). \par

Even if the first two coefficients of $\beta$ functions in different schemes coincide,
the running coupling constants - at the same ``renormalization point'' $\mu \sim k$,
but in different renormalization schemes - are generally different. If the
renormalization group equation for a running coupling constant is solved explicitly,
the description of its scale dependence necessitates the introduction of an (invariant)
parameter $\Lambda$. The difference between running coupling constants in different
renormalization schemes can then be described in terms of different $\Lambda$
parameters. \par

In the case of a perturbative treatment of gauge theories, much use has been made of a
scheme based on dimensional regularization and a modified form of minimal subtraction,
called the $\overline{MS}$ scheme \cite{13r}. Since it has become clear, how coupling
constants or $\Lambda$ parameters of different renormalization schemes can be related
explicitly \cite{12r}, the $\Lambda$ parameters of most schemes have been expressed in
terms of $\Lambda_{\overline{MS}}$. The purpose of the present paper is the
derivation of the relation between the $\Lambda$ parameter within an ERGE scheme
(defined more precisely below) and $\Lambda_{\overline{MS}}$. We will proceed as 
follows~: first, we briefly repeat the relation between the finite part of
the counter terms with different renormalization schemes, and the different $\Lambda$
parameters, following ref. \cite{12r}. Second, we will clarify, that a particular
definition of the running gauge coupling, within the ERGE approach,
corresponds - implicitly - to a particular choice of the finite parts of the counter
terms. In the next chapter we will compute, within dimensional
regularization and for a particularly simple definition of the running gauge coupling, 
these finite parts of the counter terms explictly, which
will allow us to relate a $\Lambda$ parameter denoted by $\Lambda_{ERGE}$ to
$\Lambda_{\overline{MS}}$. \par

Let us now briefly review the role of the finite part of the counter terms. We will
denote by $g_{bare}$ the bare coupling constant, by $g_a$ the renormalized coupling
constant in the renormalization scheme $a$, and by $Z_i^a$ the
renormalization constants in the scheme $a$. Generally, within dimensional
regularization and either a mass independent renormalization scheme or after
choosing a renormalization point $M = \mu$, the renormalization constants are of
the form 

\beq
Z_i^a = \left [ 1 + g_a^2 \left ( {b_i \over \varepsilon} + c_{a,i} \right ) + {\cal
O}(\varepsilon ) + {\cal O} (g_a^4) \right ] \label{2.7e}
\eeq     

\noi with $\varepsilon = 4 - d$, $b_i$ and $c_{a,i}$ are finite numerical coefficients~;
note that the $b_i$ are scheme independent. (In gauge theories, both $b_i$ and $c_{a,i}$
may, however, depend on the gauge parameter $\alpha$). \par

The relation between the bare and the renormalized coupling is generally of the form

\beq
g_{bare} = \mu^{{\varepsilon \over 2}} g_a \cdot \prod_i (Z_i^a)^{p_i}
\label{2.8e}
\eeq

\noi where the powers $p_i$ depend on the vertex used to define the renormalized
coupling, and $\mu$ is the scale introduced by the dimensionful gauge coupling constant
in $d = 4 - \varepsilon$ dimensions. \par

Within a different renormalization scheme $b$, both eqs. (\ref{2.7e}) and (\ref{2.8e})
hold with the indices $a$ replaced by indices $b$. Then, one can use the scheme
independence of $g_{bare}$ in order to derive a relation between the different
renormalized couplings~:

\beq
g_b = g_a \prod_i \left ( {Z_{i}^a \over Z_i^b} \right )^{p_i} \ \ \ .
\label{2.9e}
\eeq  

To lowest nontrivial order within an expansion in powers of the renormalized coupling,
one can neglect the difference between $g_a$ and $g_b$ within the renormalization
constants $Z_i^a$ and $Z_i^b$, and one obtains a relation between $g_a$ and $g_b$ of the
form

\beq
g_b = g_a \left ( 1 + g_a^2 \sum_i p_i (c_{a,i} - c_{b, i}) + {\cal O}(g_a^4) \right ) \
\ \ . \label{2.10e}
\eeq

This relation can be translated into a relation among the different $\Lambda$
parameters. First, from the $\mu$ independence of $g_{bare}$ in eq. (\ref{2.8e}) one
obtains the renormalization group equation

\beq
\mu^2 {dg_a^2 \over d\mu^2} = - \beta_0 \ g_a^4 - \beta_1 \ g_a^6 + {\cal O}(g_a^8) \ \ \
. \label{2.11e}
\eeq

\noi This equation can be solved in powers of $[\ell n (\mu^2/\Lambda_a^2)]^{-1}$, where
$\Lambda_a$ is a $\mu$ independent dimensionful parameter~:

\beq
g_a^2 = {1 \over \beta_0 \ell n (\mu^2/\Lambda_a^2)} - {\beta_1 \ell n \ \ell n
(\mu^2/\Lambda_a^2) \over \beta_0^3 \ell n^2 (\mu^2 /\Lambda_a^2)} + {\cal O} (\ell
n^{-3} \left ( \mu^2/\Lambda_a^2) \right )  \ \ \ . \label{2.12e} \eeq
 
From (\ref{2.12e}) and eq. (\ref{2.10e}) one then obtains

\beq
\ell n \left ( {\Lambda_b^2 \over \Lambda_a^2} \right ) = {2 \over \beta_0} \sum_i p_i
(c_{a,i} - c_{b,i}) + {\cal O}(g_a^2) \ \ \ . \label{2.13e}
\eeq

\noi The leading term on the r.h.s. of (\ref{2.13e}), which does not vanish for $\mu \to
\infty$ ($g_a \to 0$), can thus be found from the finite parts of the renormalization
constants (\ref{2.7e}) evaluated to one-loop order. \par

The next step consists in identifying the renormalization condition, or the choice of
the finite part of the renormalization constants, which is implicit in the ERGE
approach. To this end we consider an iterative solution of the ERGE (\ref{2.6e}), to
first order, expressed in terms of a bare $k$ independent action $\Gamma_{bare}(\varphi
)$ and a dimensionally regularized momentum integration~:

\beq
\Gamma_k (\varphi ) = \left . \left [ \Gamma_{bare}(\varphi ) + {\mu^{\varepsilon} \over
2} \int {d^{4- \varepsilon} q \over (2 \pi )^{4 - \varepsilon}} \ell n \left ( {\delta^2
\Gamma_{bare}(\varphi ) \over \delta \varphi (q) \delta \varphi (- q)} + R_k(q^2) \right
) \right ] \right |_{\varepsilon \to 0} \ . \label{2.14e} \eeq

Note that $\Gamma_{bare}(\varphi )$ has to contain $\varepsilon$ dependent counter
terms in order to render the r.h.s. of eq. (\ref{2.14e}) finite for $\varepsilon \to
0$, but that the $k$ derivative of the second term on the r.h.s. of eq. (\ref{2.14e})
contains an ultraviolet finite momentum integration, which allows the limit 
$\varepsilon \to 0$ of $\partial_k \Gamma_k$. \par

The finite parts of the counter terms in $\Gamma_{bare}(\varphi )$ can be chosen such
that $\Gamma_k(\varphi )$ satisfies a desired boundary condition. This boundary
condition has to coincide with the starting point of the integration of the ERGEs,
namely a ``high energy'' effective action $\Gamma_{^{-}\hskip - 2.5 truemm{k}}(\varphi )$
for some large scale $^{-}\hskip - 2.5 truemm{k}$. Using eq. (\ref{2.14e}) we can
express $\Gamma_{bare}(\varphi)$ explicitly in terms of $\Gamma_{^{-}\hskip - 2.5
truemm{k}}(\varphi )$ since, to first order, we can replace the action in the argument
of the logarithm by $\Gamma_{^{-}\hskip - 2.5 truemm{k}}$~:

\beq
\Gamma_{bare}(\varphi ) = \Gamma_{^{-}\hskip - 2.5 truemm{k}}(\varphi ) -
{\mu^{\varepsilon} \over 2} \int {d^{4 - \varepsilon} q \over (2 \pi )^{4- \varepsilon}}
\ell n \left ( {\delta^2 \Gamma_{^{-}\hskip - 2.5 truemm{k}}(\varphi ) \over \delta
\varphi (q) \delta \varphi (- q)} + R_{^{-}\hskip - 2.5 truemm{k}} (q^2) \right ) \ .
\label{2.15e} \eeq

\noi Given an ansatz for $\Gamma_{^{-}\hskip - 2.5 truemm{k}} (\varphi )$, and
expanding both sides of (\ref{2.15e}) in powers of fields, we can now construct the
renormalization constants $Z_i^{ERGE}$ explicitly. A priori they will depend on the
scale $^{-}\hskip - 2.5 truemm{k}$, appearing as infrared cutoff on the r.h. side on
(\ref{2.15e}). Only after identifying $^{-}\hskip - 2.5 truemm{k}$ with $\mu$ the
renormalization constants will be of the form (\ref{2.7e}). Then we can read off the
finite coefficients $c_{ERGE, i}$. Since the corresponding coefficients in the
$\overline{MS}$ scheme $c_{\overline{MS}, i}$ are already known \cite{12r}, we are
subsequently able to relate the parameters $\Lambda_{ERGE}$ and
$\Lambda_{\overline{MS}}$ using eq. (\ref{2.13e}). \par

Note that, although we encountered divergent expressions (for $\varepsilon \to 0$) in
the form of eqs. (\ref{2.14e}) and (\ref{2.15e}), ultraviolet divergences are absent if
we express $\Gamma_k(\varphi )$ in terms of $\Gamma_{^{-}\hskip - 2.5 truemm{k}}(\varphi
)$. Since this all that matters in the ERGE approach, there is, in principle, no need to
formulate the ERGEs in $d = 4 - \varepsilon$ dimensions. It was only our desire to
relate $\Gamma_{^{-}\hskip - 2.5 truemm{k}}(\varphi )$ to some bare action within
dimensional regularization, in order to make the relation with the $\overline{MS}$ scheme
explicit, which forced us away from four dimensions. In the next chapter we will proceed
towards the explicit calculation of the renormalization constants $Z_i^{ERGE}$ in
Yang-Mills theories.

\mysection{Renormalization constants for Yang-Mills theories in the ERGE scheme}

Let us first present our conventions for the classical action of a $SU(N)$ Yang-Mills
theory in four-dimensional Euclidean space-time, with the usual gauge fixing and ghost
parts included, and with external sources $K_{\mu}^a$, $L^a$ and $\bar{L}^a$ coupled to
the BRST variations of the fields $A_{\mu}^a$, $c^a$ and $\bar{c}^a$ respectively~:

\bea
S &=& \int d^4 x \left \{ {1 \over 4} F_{\mu \nu}^a \ F_{\mu \nu}^a + {1 \over 2 \alpha}
\partial_{\mu} A_{\mu}^a \partial_{\nu} A_{\nu}^a + 
+ \partial_{\mu} \bar{c}^a \left ( \partial_{\mu} c^a + g f^a_{\ bc} A_{\mu}^a c^c
\right ) \right . \nn \\
&-& \left . K_{\mu}^a \left ( \partial_{\mu} c^a + g f_{\ bc}^a A_{\mu}^a c^c \right ) 
-  {1 \over 2} g L^a \ f_{\ bc}^a c^b c^c + {1 \over \alpha} \bar{L}^a
\partial_{\mu} A_{\mu}^a \right \}
\label{3.1e} 
\eea

\noi with 

\beq
F_{\mu \nu}^a = \partial_{\mu} A_{\nu}^a - \partial_{\nu} A_{\mu}^a + g \ f_{\ bc}^a
A_{\mu}^b A_{\nu}^c \ \ \ . \label{3.2e}
\eeq 

The ERGE approach demands the presence of infrared cutoff terms, cf. eqs. (\ref{2.1e})
and (\ref{2.3ea}), for the gluon and ghost propagators~:

\beq
\Delta S_k = {1 \over 2} \left ( A_{\mu}^a , R_{k, \mu \nu} (p^2) A_{\nu}^a \right ) +
\left ( \bar{c}^a , \widetilde{R}_k(p^2) c^a \right ) \label{3.3e}
\eeq

\noi Explicit expressions for $R_{k, \mu \nu }$ and $\widetilde{R}_k$ will be given
below. The fact that the presence of $\Delta S_k$ in eq. (\ref{2.1e}) breaks gauge or
BRST invariance can be translated into a modification of the Slavnov-Taylor identities,
which have to be satisfied by the effective action $\Gamma_k(A_{\mu}^a , c^a,
\bar{c}^a)$ constructed along eqs. (\ref{2.1e}) and (\ref{2.5e}) \citd{8r}{9r}. They
read

\bea
&&\left . \int {d^4p \over (2 \pi )^4} \left \{ {\delta \widehat{\Gamma}_k \over \delta
K_{\mu}^a (-p)} {\delta \widehat{\Gamma}_k \over \delta A_{\mu}^a(p)} - {\delta
\widehat{\Gamma}_k \over \delta L^a(-p)} {\delta \widehat{\Gamma}_k \over \delta c^a
(p)} - {\delta \widehat{\Gamma}_k \over \delta \bar{L}^a (-p)} {\delta
\widehat{\Gamma}_k \over \delta \bar{c}^a (p)} \right \} \right |_{\bar{L} = 0} \nn \\
&&= \int {d^4 p \over ( 2 \pi )^4} \sum_B \left [ R_{k, \mu \nu} (p^2) {\delta^2
\widehat{\Gamma}_k \over \delta K_{\nu}^a(- p) \delta \varphi^B}
\left ( \Gamma_k^{(2)} \right )^{-1}_{\bar{\varphi}^B, A_{\mu}^a (p)} \right . \nn \\ 
&&\left . \left . - \widetilde{R}_k(p^2) {\delta^2 \widehat{\Gamma}_k \over \delta L^a(-p)
\delta \varphi^B} \left ( \Gamma_k^{(2)} \right )^{-1}_{\bar{\varphi}^B , c^a (p)} -
\widetilde{R}_k(p^2) {\delta^2 \widehat{\Gamma}_k \over \delta \bar{L}^a (-p) \delta
\bar{\varphi}^B} \left ( \Gamma_k^{(2)} \right )^{-1}_{\varphi^B, \bar{c}^a (p)} \right
] \right |_{\bar{L} = 0}  \label{3.4e} \eea

\noi where the sum over $B$ runs over the fields $\varphi^B = \{A_{\mu}^a(p), c^a(p),
\bar{c}^a(p)\}$ and $\bar{\varphi}^B = \{ A_{\mu}^a (-p)$, $- \bar{c}^a (-p)$, $c^a
(-p)\}$. $\Gamma_k$ and $\widehat{\Gamma}_k$ are related as in eq. (\ref{2.5e}). $\left
( \Gamma_k^{(2)} \right )^{-1}_{\bar{\varphi}^B \varphi^C}$ denotes the
$(\bar{\varphi}^B, \varphi^C)$ component of the inverse of the matrix $\Gamma_k^{(2)} =
\delta^2 \Gamma_k/\delta \bar{\varphi} \delta \varphi$ of second derivatives of
$\Gamma_k$ with respect to the field $\varphi^B$ and $\bar{\varphi}^B$. \par

The analog of the ERGE (\ref{2.6e}) reads now
\bea
\partial_k \widehat{\Gamma}_k &=& \int {d^4p \over (2 \pi )^4} \left \{ {1 \over 2}
\partial_k R_{k, \mu \nu} (p^2) \left ( \Gamma_k^{(2)} \right )^{-1}_{A_{\nu}^a (-p),
A_{\mu}^a (p)} 
%\right . \nn \\ &-& \left . 
- \partial_k \widetilde{R}_k (p^2) \left ( \Gamma_k^{(2)} \right )^{-1}_{-
\bar{c}^a(- p) , c^a (p)} \right \} \ \ \ .  \label{3.5e} \eea

In \cite{8r} it has been shown that once (\ref{3.4e}) is satisfied for some scale $k =$
$^{-}\hskip - 2.5 truemm{k}$, it will be satisfied by any $\widehat{\Gamma}_k$ provided
$\widehat{\Gamma}_k$ is obtained from $\widehat{\Gamma}_{^{-}\hskip - 2.5 truemm{k}}$
by integrating the ERGE (\ref{3.5e}). In particular $\widehat{\Gamma}_{k=0}$ will
satisfy the standard Slavnov-Taylor identity, eq. (\ref{3.4e}) with a vanishing
right-hand side, if the infrared cutoff functions $R_{k, \mu \nu}$ and
$\widetilde{R}_k$ vanish for $k \to 0$. \par

Because of the need to satisfy eq. (\ref{3.4e}) for $k =$ $^{-}\hskip - 2.5 truemm{k}$,
we are actually not free to choose the starting point $\Gamma_{^{-}\hskip - 2.5
truemm{k}}$ of the integration of the ERGEs as we wish~; in particular it cannot be
identified with the classical action (\ref{3.1e}). Perturbatively, however, a
consistent starting point $\Gamma_{^{-}\hskip - 2.5 truemm{k}}$ can be constructed
by solving (\ref{3.4e}) iteratively around the classical action (\ref{3.1e}), which
satisfies (\ref{3.4e}) with a vanishing right-hand side. Alternatively, to
first order in the coupling, a consistent starting point $\Gamma_{^{-}\hskip - 2.5
truemm{k}}$ can be constructed from eq. (\ref{2.14e})
for $k =$ $^{-}\hskip - 2.5 truemm{k}$, where $\Gamma_{bare}$ is of the form of the
classical action (\ref{3.1e}) supplemented with local counter terms. \par

Our task is, however, the construction of local counter terms from eq. (\ref{2.15e}),
using a (consistent) starting point action $\Gamma_{^{-}\hskip - 2.5 truemm{k}}$ as
input. First we emphasize a feature of the ERGE approach, which is actually well known
from momentum subtraction schemes~: since the infrared cutoff $^{-}\hskip - 2.5
truemm{k}$ breaks the BRST invariance (as, e.g., an off-shell renormalization condition),
the precise definition of the renormalization scheme (the finite parts of
the renormalization constants) depends on the vertex used to define the coupling
constant. \par

Here we will proceed by using the ghost gluon vertex for this definition. The terms in
the bare action, which involve the renormalization constants $Z_A$, $Z_c$, and
$Z_{Ac\bar{c}}$ required to relate the renormalized coupling $g_{ERGE}$ to the bare
coupling $g_{bare}$ in this case, are as follows~:

\bea
\Gamma_{bare} &=& {1 \over 2} \int_{p_1, p_2} A_{\mu}^a(p_1) \left \{ Z_A (p_1^2
\delta_{\mu \nu} - p_{1 \mu} p_{1\nu}) + {p_{1\mu} p_{1\nu} \over \alpha} \right \}
A_{\nu}^a (p_2) \nn \\
&+& \int_{p_1, p_2} \bar{c}^a(p_1) Z_c p_1^2 c^a(p_2) \nn \\
&+& i \ g_{ERGE} f_{\ bc}^a \int_{p_1, p_2, p_3} \bar{c}^a(p_1) Z_{Ac \bar{c}} \ p_{1,
\mu} \ A_{\mu}^b(p_2) \ c^c(p_3) + \cdots   \label{3.6e} \eea  

\noi with

\beq
\int_{p_1, \cdots , p_n} \equiv \int \prod^n_{i=1} \left ( {d^d p_i \over ( 2 \pi )^d}
\right ) \cdot (2 \pi )^d \delta^d \left ( \sum_{i=1}^n p_i \right ) \ \ \ . \label{3.7e}
\eeq
 
\noi The relation between $g_{ERGE}$ and $g_{bare}$ is then

\beq
g_{ERGE} = {Z_A^{1/2} Z_c \over Z_{Ac \bar{c}}} g_{bare} \ \ \ .
\label{3.8e}
\eeq

\noi In order to compute the renormalization constants $Z_A$, $Z_c$ and $Z_{Ac\bar{c}}$
from the analog of eq. (\ref{2.15e}) for Yang-Mills theories, we have to consider the
terms of ${\cal O}(A^2)$, ${\cal O}(c \bar{c})$ and ${\cal O}(Ac \bar{c})$,
respectively. Now note that the corresponding terms of eq. (\ref{2.15e}) relate momentum
dependent vertex functions of $\Gamma_{bare}$ and $\Gamma_{^{-}\hskip - 2.5
truemm{k}}$~: if we require that $\Gamma_{bare}$ is of the form of the classical action
(\ref{3.1e}) supplemented with local counter terms (and satisfies the classical
Slavnov-Taylor identity), $\Gamma_{^{-}\hskip - 2.5 truemm{k}}$ necessarily has to
contain nontrivial momentum dependent vertex functions. These nontrivial momentum
dependent vertex functions correspond to the one-loop improved starting point action
employed in ref. \cite{9r}. \par

As a result we have to decide, which values of the external momenta we use, in order to
fix the definition of the gauge coupling and thus the 
renormalization scheme completely. The simplest choice is, of course, vanishing
external momenta. (For external momenta $p^2 =$ $^{-}\hskip - 2.5 truemm{k}^2$, e.g., we
would obtain different finite parts of the counter terms). \par

The diagrams which contribute to $Z_A$, $Z_c$ and $Z_{Ac \bar{c}}$ are shown in figs. 1,
2 and 3, respectively. Of course we have to take appropriate derivatives with respect to
the external momenta before setting the external momenta to zero, in order to reproduce
the tensor structures in eq. (\ref{3.6e}). For completeness, we took in $Z_A$ the
presence of $n_F$ massless fermions in the fundamental representation of $SU(N)$ into
account~; their propagator is supposed to be cutoff in the infrared due to a
contribution to $\Delta S_k$, eq. (\ref{3.3e}), of the form $(\bar{\psi}^a, \not p
R_k^{\psi}(p^2) \psi ^a)$. \par

Next we have to specify the infrared cutoff terms $R_{k, \mu \nu}$, $\widetilde{R}_k$
and $R_k^{\psi}$ for the gluon, ghost and fermion fields. In principle these terms could
depend on the $k$-dependent parameters of $\Gamma_k$ \cite{9r} (in order to ensure the
absence of unwanted poles in the propagators), but to the presently required order a
simple choice analogous to eq. (\ref{2.3ea}) is sufficient~:

\bminiG{EDh}
\widetilde{R}_k(p^2) = p^2 {e^{- {p^2 \over k^2}} \over 1 - e^{- {p^2 \over k^2}}}
\label{3.9ea}
\eeeq
\beeq
R_{k, \mu \nu}(p) = \widetilde{R}_k (p^2) \left \{ \delta_{\mu \nu} - \left ( 1 - {1
\over \alpha} \right ) {p_{\mu} p_{\nu} \over p^2} \right \} \label{3.9eb}
\eeeq
\beeq
R_k^{\psi} (p^2) = {e^{-{p^2 \over k^2}} \over 1 - e^{-{k^2 \over p^2}}} \ \ \ . 
\label{3.9ec}
\emini

\noi Finally we fix the gauge~; the Landau gauge $\alpha = 0$ turns out to be
particularly convenient, because in this gauge the diagrams in fig. 3 vanish (for
vanishing external momenta). \par

We then obtain~:

\bminiG{q-def}
Z_A^{ERGE} = 1 + {g^2 \over 16 \pi^2} \left [ {2 \over \varepsilon} \left ( {13 \over
6} N - {2 \over 3} n_F \right ) + N \left ( {13 \over 6} \ell n (4 \pi ) + {229 \over 96}
\right ) - n_F \left ( {2 \over 3} \ell n (4 \pi ) + {1 \over 36} \right ) \right ] 
\label{3.10ea} \eeeq
\beeq 
Z_c^{ERGE} = 1 + {g^2 \over 16 \pi^2} \left [ {2 \over \varepsilon}
\left ( {3 \over 4} N \right ) + N \left ( {3 \over 4} \ell n (4 \pi ) + {1 \over 4}
\right ) \right ] \label{3.10eb} \eeeq
\beeq 
Z_{Ac \bar{c}}^{ERGE} = 1 \ \ \ . \label{3.10ec} 
\emini

The corresponding renormalization constants in the $\overline{MS}$ scheme, one the other
hand, are as follows \cite{12r} (note that $d = 4 + \varepsilon$ in \cite{12r}, whereas
$d = 4 - \varepsilon$ here)~:

\bminiG{q-def1}
Z_A^{\overline{MS}} = 1 + {g^2 \over 16 \pi^2} \left [ {13 \over 6} N \left ( {2
\over \varepsilon} + \ell n (4 \pi ) - \gamma_E \right ) - {2 \over 3} n_F \left ( {2
\over \varepsilon} + \ell n (4 \pi ) - \gamma_E \right ) \right ]  \label{3.11ea} \eeeq
\beeq  Z_c^{\overline{MS}} = 1 + {g^2 \over 16 \pi^2} \left [ {3 \over 4} N \left ( {2
\over \varepsilon} + \ell n (4 \pi ) - \gamma_E  \right ) \right ] \label{3.11eb} \eeeq
\beeq 
Z_{Ac \bar{c}}^{\overline{MS}} = 1  \label{3.11ec} 
\emini

\noi where $\gamma_E$ is Euler's constant~: $\gamma_E = 0,577216$ ... \par

{From} eq. (\ref{3.8e}), and the analogs of eqs. (\ref{2.7e}) and
(\ref{2.10e}), we thus find

\beq
g_{ERGE} = g_{\overline{MS}} \left ( 1 + g^2_{\overline{MS}} \left [ N \left ( {11 \over
6} \gamma_E + {277 \over 192} \right )  - n_F \left ( {1 \over 3} \gamma_E + {1 \over
72} \right ) \right ] + {\cal O} \left ( g^4_{\overline{MS}} \right ) \right )
\label{3.12e} \eeq

\noi or, from eq. (\ref{2.13e}) and with $\beta_0 = \left ( \displaystyle{{11 \over 3}} N
- \displaystyle{{2 \over 3}} n_F \right )/16 \pi^2$,

\beq
\ell n \left ( {\Lambda^2_{ERGE} \over \Lambda^2_{\overline{MS}}} \right ) = {N \left (
{11 \over 3} \gamma_E + {277 \over 96} \right ) - n_F \left ( {2 \over 3} \gamma_E + {1
\over 36} \right ) \over {11 \over 3} N - {2 \over 3} n_F} + {\cal O} (g^2) \ \ .
\label{3.13e} \eeq 

\noi This is our main result~; numerically it leads, for $n_F = 0$ and $N$ arbitrary
(pure Yang-Mills), to $\Lambda_{ERGE} \sim 2 \Lambda_{\overline{MS}}$. \par

In the remaining part of this section we will briefly discuss, how this result may be
applied within the ERGE formalism. Let us note again, that the starting point in the
ERGE formalism is a high energy effective action $\Gamma_{^{-}\hskip - 2.5 truemm{k}}$.
The only free parameter in $\Gamma_{^{-}\hskip - 2.5 truemm{k}}$ is a coupling constant
$g_{^{-}\hskip - 2.5 truemm{k}}$. Due to the one-loop improvement of $\Gamma_{^{-}\hskip
- 2.5 truemm{k}}$ required to satisfy eqs. (\ref{2.14e}) and (\ref{2.15e}) with a local
bare action, or, equivalently, the modified Slavnov-Taylor identity (\ref{3.4e}), the
precise definition of $g_{^{-}\hskip - 2.5 truemm{k}}$ will depend on the vertex function
used, on the external moment and the gauge. Here we confine ourselves to the ghost gluon
vertex at vanishing external momenta and the Landau gauge. \par

As a result of the integration of the ERGEs one can obtain dimensionful nonperturbative
quantities like hadron masses or the slope $\sigma$ of a confining potential. For
dimensional reasons quantities like $\sigma$ are necessarily given in units of the
ultraviolet starting scale $^{-}\hskip - 2.5 truemm{k}$, on the other hand they have to
depend on $^{-}\hskip - 2.5 truemm{k}$ and $g_{^{-}\hskip - 2.5 truemm{k}}$ such that
their total derivative with respect to $^{-}\hskip - 2.5 truemm{k}$ vanishes~:

\beq
\sigma^2 = \gamma ^{-}\hskip - 2.5 truemm{k}^2 e^{- {1 \over \beta_0 g^2_{^{-}\hskip - 2.5
truemm{k}}}} \left ( \beta_0 \ g^2_{^{-}\hskip - 2.5 truemm{k}} \right )^{- {\beta_1
\over \beta_0^2}} \left ( 1 + {\cal O} \left ( g^2_{^{-}\hskip - 2.5 truemm{k}} \right )
\right )   \label{3.14e} \eeq

\noi with $\gamma$ a $^{-}\hskip - 2.5 truemm{k}$ independent constant. Eq.
(\ref{3.14e}) can and has to be checked within the ERGE formalism, for $g_{^{-}\hskip -
2.5 truemm{k}}$ small enough. Most importantly, it requires that $g_{^{-}\hskip - 2.5
truemm{k}}$ runs with $^{-}\hskip - 2.5 truemm{k}$ according to the renormalization
group equation (\ref{2.11e}) with the two-loop term $\beta_1$ included. Only then it
becomes possible to extract from eq. (\ref{3.14e}) the quantity $\gamma$ in the limit
$g_{^{-}\hskip - 2.5 truemm{k}} \to 0$. \par

Using eq. (\ref{2.12e}) with $\mu^2 =$ $^{-}\hskip - 2.5 truemm{k}^2$ and $\Lambda_a =
\Lambda_{ERGE}$, eq. (\ref{3.14e}) becomes, as it should,

\beq
\sigma^2 = \gamma \Lambda_{ERGE}^2 \left ( 1 + {\cal O} \left ( g_{^{-}\hskip - 2.5
truemm{k}}^2 \right ) \right ) \ \ \ .  \label{3.15e}
\eeq

\noi Hence, having determined $\gamma$ within the ERGE approach and having compared
$\sigma$ with some measured quantity, one is able to determine $\Lambda_{ERGE}$ in,
say, MeV. At this stage we can apply our result, eq. (\ref{3.13e}), and obtain
$\Lambda_{\overline{MS}}$ in MeV. This information would certainly be highly welcome.
\par

A first effort in this direction has been made in ref. \cite{9r}. There the ERGEs for
the gluon and ghost propagators in pure SU(3) Yang-Mills theory were integrated, within
a certain approximation, with the aim to study the potential between heavy quarks. The
result was indeed a confining form of the potential, which allows the determination of
a phenomenologically known quantity like $\sigma$ in eqs. (\ref{3.14e}) and
(\ref{3.15e}). In addition, a one-loop improvement of the action $\Gamma_{^{-}\hskip -
2.5 truemm{k}}$ at the starting point was used, which renders the definition of the
coupling $g_{^{-}\hskip - 2.5 truemm{k}}$ free from the am\-bi\-gui\-ties discussed above.
However, the approximations performed on the r.h. sides of the ERGEs in \cite{9r} (in
order to turn them into a low-dimensional closed system of differential equations)
were too crude to reproduce correctly the two-loop coefficient $\beta_1$ of the
$\beta$-function for $g_{^{-}\hskip - 2.5 truemm{k}}$. Therefore eq. (\ref{3.14e}) was
not satisfied to the required order, and a parameter $\gamma$ as in eq. (\ref{3.15e})
could not be extracted. It should be clear that the determination of
$\Lambda_{\overline{MS}}$ in the ERGE approach allows only for approximations, which
reproduce correctly the two-loop $\beta$-function.

\mysection{Conclusions}

In the present paper we have discussed the relation between the ERGE approach and the
standard renormalization procedure. In particular we have shown, that a particular
choice of the starting point action $\Gamma_{^{-}\hskip - 2.5 truemm{k}}$, within the
ERGE approach, corresponds implicitly to a particular choice of the renormalization
condition. Our aim was to establish the explicit relation eq. (\ref{3.13e}) between the
$\Lambda$ parameter of the ERGE approach (in the case of a particularly convenient
definition of the coupling $g_k^{ERGE}$, and for a given choice of the infrared
regulators $R_k(p^2)$) and $\Lambda_{\overline{MS}}$ for SU(N) gauge theories with $n_F$
massless quarks. \par

First we had to clarify, however, that the starting point action $\Gamma_{^{-}\hskip - 2.5
truemm{k}}$ in the ERGE approach has to be of the form of a one-loop improved action,
in order to render the relation between the coupling constants in the ERGE approach
and the $\overline{MS}$ scheme free from ambiguities~: if one would naively choose
$\Gamma_{^{-}\hskip - 2.5 truemm{k}}$ to be of the form of the classical action, the
definition of the coupling constant would, of course, not depend on the vertex nor on
the external momenta of the corresponding particles. The finite parts of the
counterterms, which relate $\Gamma_{^{-}\hskip - 2.5 truemm{k}}$ to $\Gamma_{bare}$,
do, however, depend on these conventions and would therefore be ambiguous. From the
one-loop relation between $\Gamma_{^{-}\hskip - 2.5 truemm{k}}$ and $\Gamma_{bare}$
(which is necessary and sufficient for the present discussion) one also sees that a
``classical'' choice of $\Gamma_{^{-}\hskip - 2.5 truemm{k}}$ corresponds, in fact, to
a nonlocal form of $\Gamma_{bare}$ which does not coincide with the bare action
implicit in the $\overline{MS}$ scheme. \par

Second, as one may have suspected, a practical application of our result in order to
determine $\Lambda_{\overline{MS}}$ from the ERGE approach requires an approximation,
which is general enough in order to reproduce the two-loop $\beta$ function. This
condition is certainly much harder to satisfy, in practice, than the use of a
one-loop improved starting point action (which has already been used in \cite{9r})
although it would be, in principle, straightforward to generalize the approach in
\cite{9r} sufficiently. In any case the present result is a prerequisite for such a
program.  \par \vskip 5 truemm

\noi {\bf\large Acknowledgement} \par

It is a pleasure to thank B. Bergerhoff and A. Weber for useful discussions.

\vspace{2 cm}

%\newpage
\def\labelenumi{[\arabic{enumi}]}
\noindent
{\bf\large References}
\ben
\item\label{1r} K. G. Wilson and I. Kogut, Phys. Rep. {\bf 12} (1974) 75~; F. Wegner,
in : Phase Transitions and Critical Phenomena, Vol. 6, eds. C. Domb and M. Green
(Academic Press, NY 1975).
\item\label{2r} J. Polchinski, Nucl. Phys. {\bf B231} (1984) 269.  
\item\label{3r} F. Wegner and A. Houghton, Phys. Rev. {\bf A8} (1973) 401~; A.
Hasenfratz and P. Hasenfratz, Nucl. Phys. {\bf B270} (1986) 687.  
\item\label{4r} S. Weinberg, in~: Proceedings of the 1976 International School of
Subnuclear Physics, Erice (ed. A. Zichichi). 
\item\label{5r} G. Keller and C. Kopper, Phys. Lett. {\bf B273} (1991) 323~; G. Keller,
Helv. Phys. Acta {\bf 66} (1993) 453.  
\item\label{6r} M. Reuter and C. Wetterich, Nucl. Phys. {\bf B417} (1994) 181, ibid.
{\bf B427} (1994) 291~; Phys. Lett. {\bf B334} (1994) 412.   
\item\label{7r} C. Becchi, in~: Elementary Particles, Field Theory and Statistical
Mechanics, eds. M. Bonini, G. Marchesini and E. Onofri, Parma University 1993~; M.
Bonini, M. D'Attanasio and G. Marchesini, Nucl. Phys. {\bf B418} (1994) 81, ibid. {\bf
B421} (1994) 429, ibid. {\bf B437} (1995) 163, Phys. Lett. {\bf B346} (1995) 87~; M.
D'Attanasio and T. Morris, Phys. Lett. {\bf B378} (1996) 213~; M. Bonini and G.
Marchesini, Phys. Lett. {\bf B389} (1996) 566~; M. Bonini, G. Marchesini and M.
Simionato, hep-th/9604114.
\item\label{8r} U. Ellwanger, Phys. Lett. {\bf B335} (1994) 364~; U. Ellwanger, M.
Hirsch and A. Weber, Z. Phys. {\bf C69} (1996) 687.   
\item\label{9r} U. Ellwanger, M. Hirsch and A. Weber, Orsay preprint 96-50,
hep-ph/9606468.  
\item\label{10r} J. L. Richardson, Phys. Lett. {\bf B82} (1979) 272. 
\item\label{11r} C. Wetterich, Phys. Lett. {\bf B301} (1993) 90.   
\item\label{12r} W. Celmaster and R. Gonsalves, Phys. Rev. {\bf D20} (1979) 1420. 
\item\label{13r} W.A. Bardeen, A. Buras, D. Duke and T. Muta, Phys. Rev. {\bf D18}
(1978) 3998.
\een 

\newpage

\begin{figure}
\centering
\includegraphics[width=2.5in]{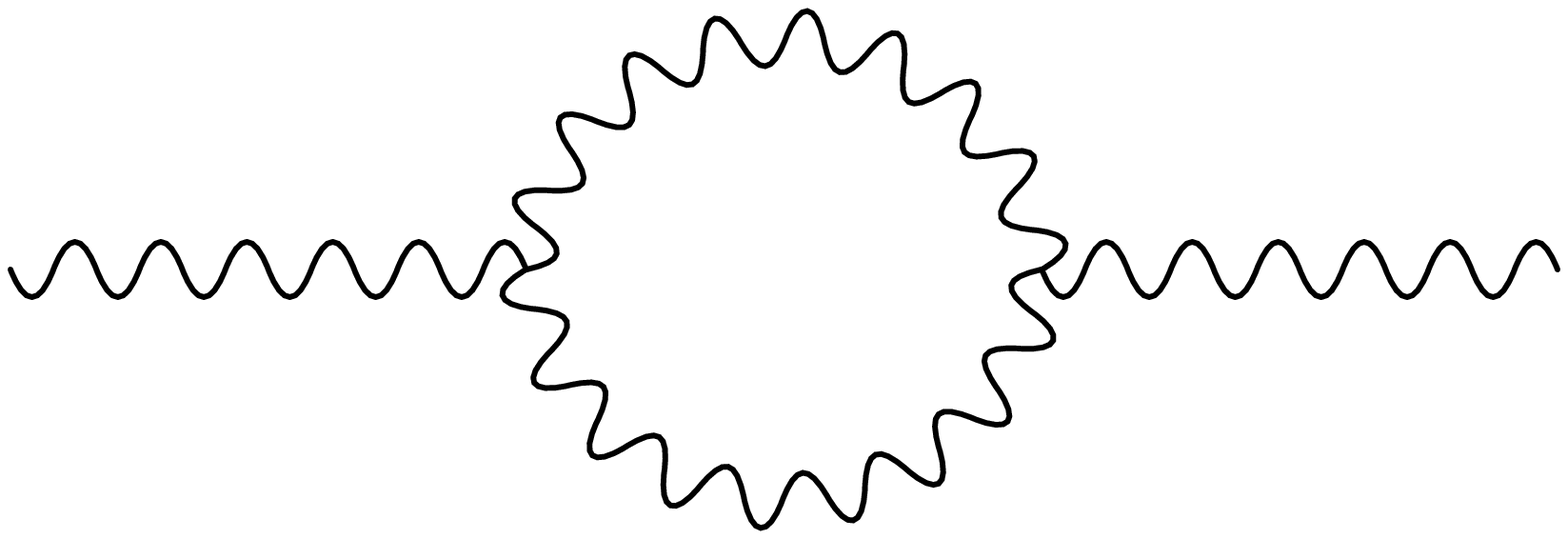}
\hspace{1in}
\includegraphics[width=2.5in]{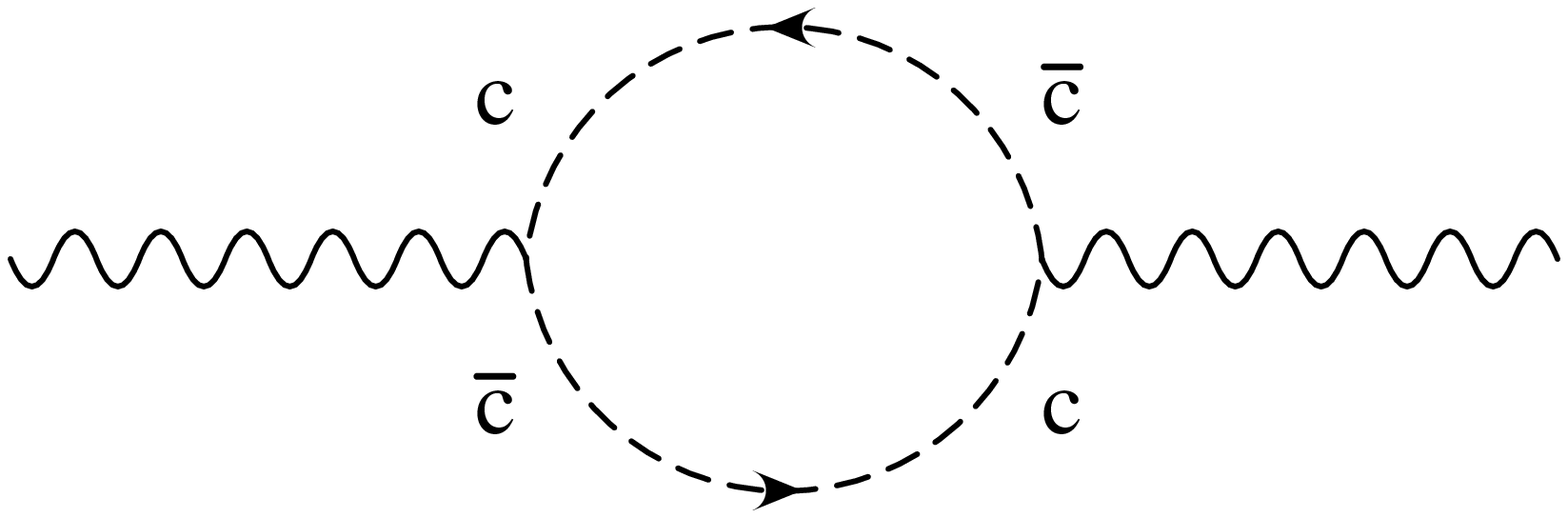}

\includegraphics[width=2.5in]{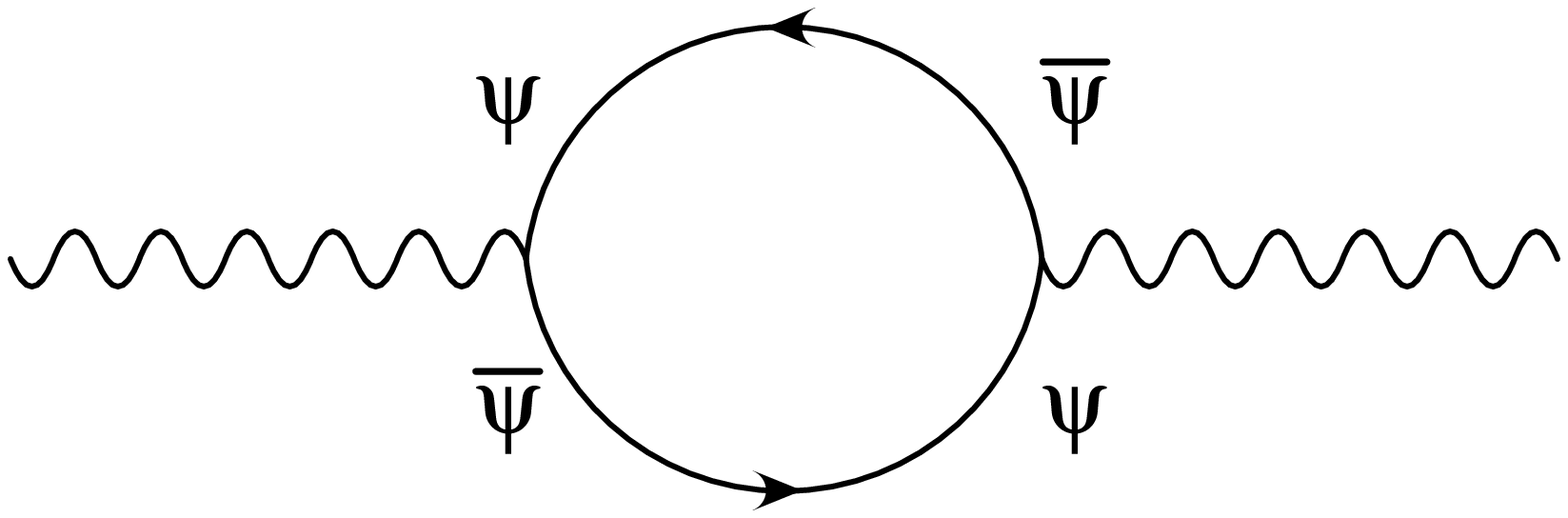}
\caption{Diagrams contributing to $Z_A$. All internal propagators are
supplemented with infrared cutoffs, cf. eqs. (2.5) and (3.9). }
\end{figure}

\begin{figure}
\centering
\includegraphics[width=2.5in]{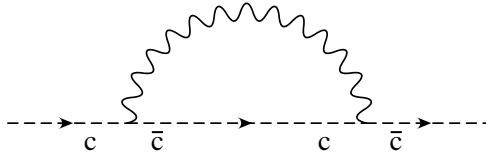}
\caption{Diagram contributing to $Z_c$.}
\end{figure}

\begin{figure}
\centering
\includegraphics[width=2.5in]{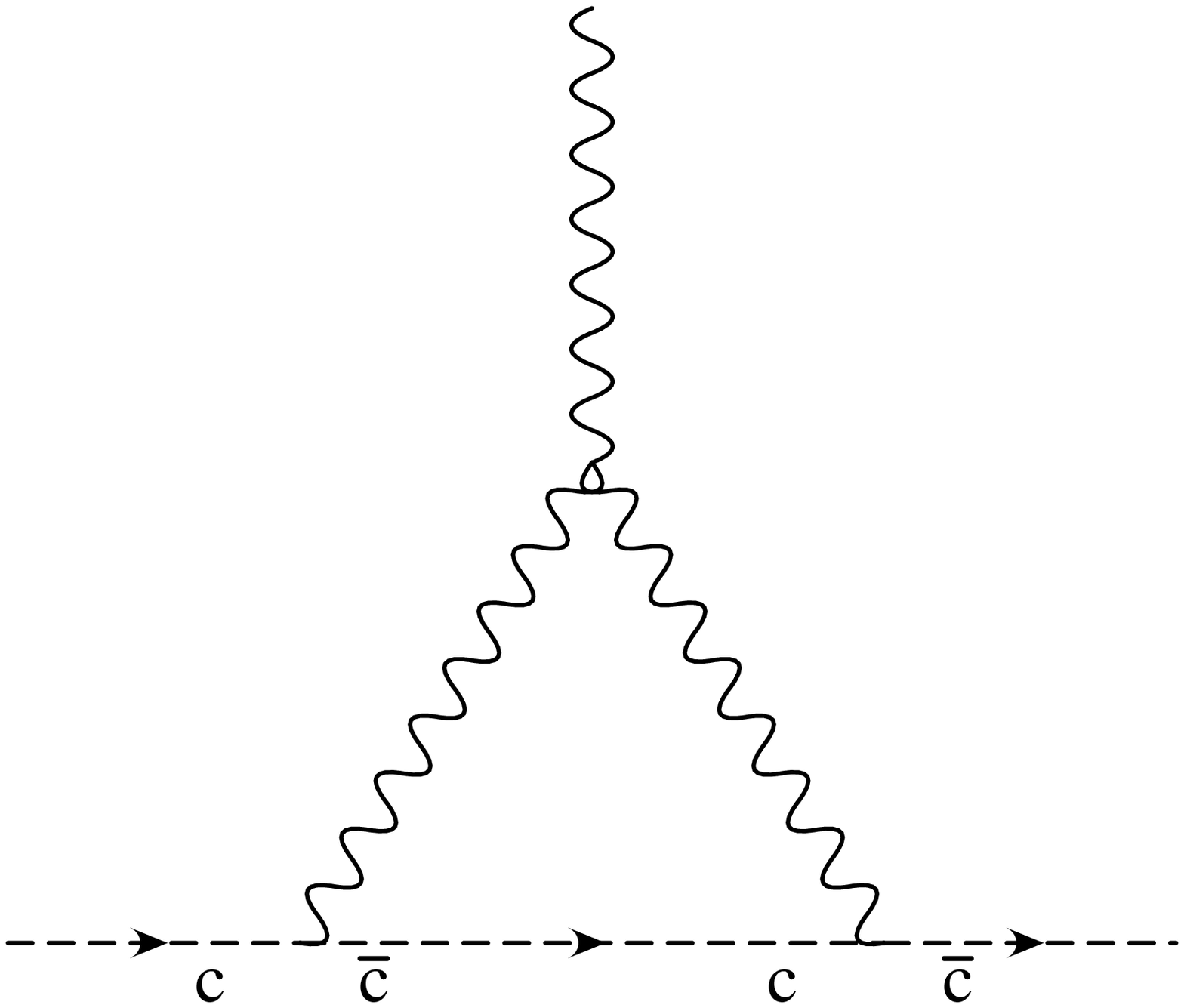}
\hspace{1in}
\includegraphics[width=2.5in]{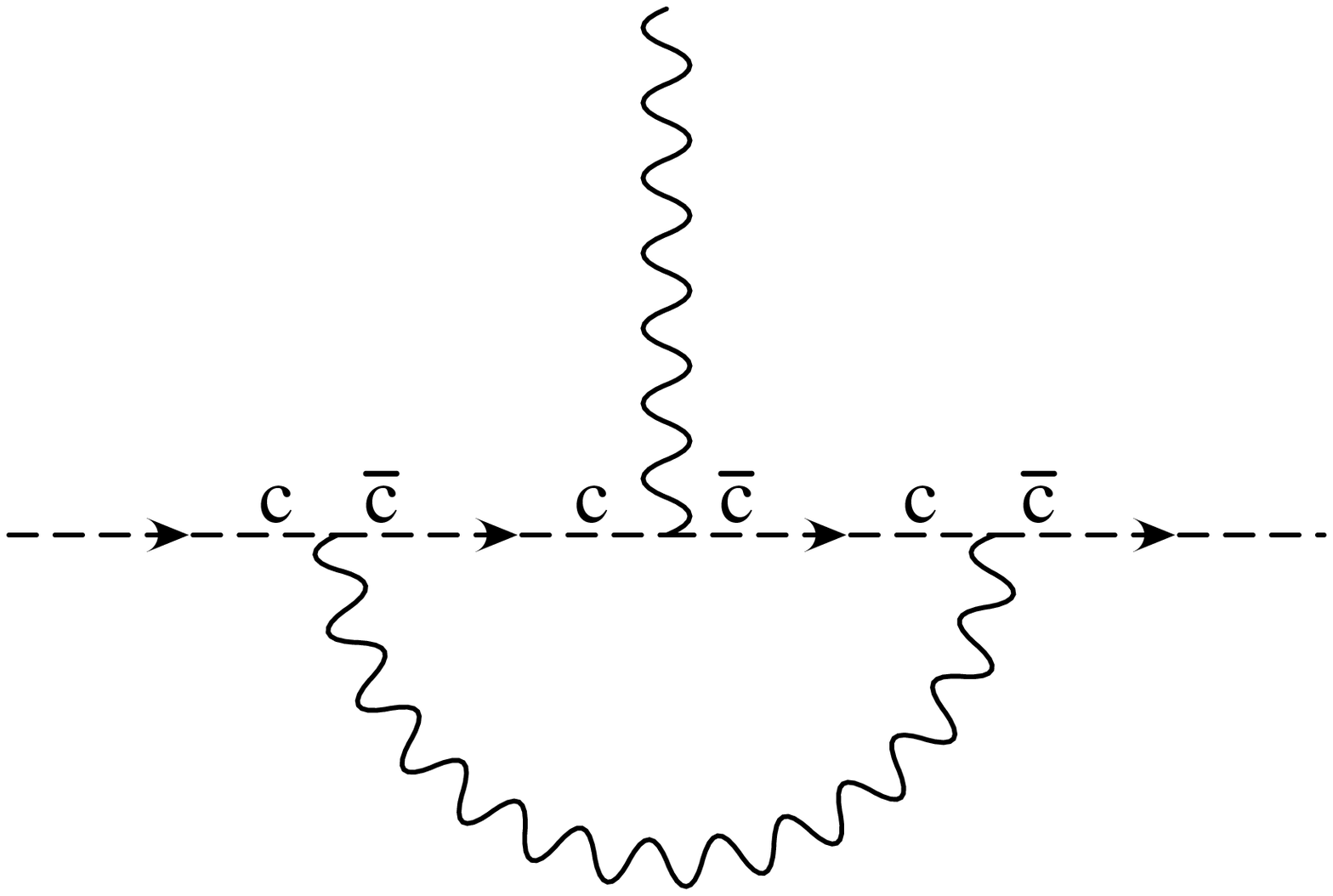}
\caption{Diagrams contributing to $Z_{Ac\bar{c}}$ (which vanish, however, in
the Landau gauge at vanishing external momenta). }
\end{figure}

\end{document}